# DFTTK: Density Functional Theory ToolKit for High-throughput Lattice Dynamics Calculations

Yi Wang, Mingqing Liao, Brandon J. Bocklund, Peng Gao, Shun-Li Shang, Hojong Kim, Allison M. Beese, Long-Qing Chen, and Zi-Kui Liu

*Department of Materials Science and Engineering, The Pennsylvania State University,*

*University Park, PA 16802, USA*

In this work, we present a software package in Python for high-throughput first-principles calculations of thermodynamic properties at finite temperatures, which we refer to as DFTTK (Density Functional Theory ToolKit). DFTTK is based on the *atomate* package and integrates our experiences in the last decades on the development of theoretical methods and computational software. It includes task submissions on all major operating systems and task execution on high-performance computing environments. The distribution of the DFTTK package comes with examples of calculations of phonon density of states, heat capacity, entropy, enthalpy, and free energy under the quasi-harmonic phonon scheme for the stoichiometric phases of Al, Ni, $Al_3Ni$, AlNi, $AlNi_3$, $Al_3Ni_4$, and $Al_3Ni_5$, and the fcc solution phases treated using the special quasirandom structures at the compositions of $Al_3Ni$, AlNi, and $AlNi_3$.

# 1 Introduction

First-principles calculations based on density functional theory (DFT) are a highly valuable tool for the materials community. By definition, the term "first-principles" represents a philosophy that the prediction is to be based on a fundamental proposition or assumption that cannot be deduced from any other proposition or assumption. DFT-based first-principles calculations rely on the Kohn-Sham equation [1,2] with the inputs from well-defined physical constants, e.g., the atomic mass and the nuclear and electronic charges, without invoking any phenomenological fitting parameters. Many large-scale public materials databases from DFT-based first-principles calculations have become available in recent years, such as *AFLOW* [3], *OQMD* [4], *The Materials Project* [5], and those recently connected by *OPTIMADE API* [6], and each has its own high throughput computation tools, typically for properties at 0 K.

For practical applications, analytical description of materials properties at finite temperatures and for multicomponent systems is needed. The corresponding modeling mechanism is usually provided by the CALPHAD (CALculation of PHAse Diagram) approach[7–9]. When not enough experimental data are available, input data from DFT-based first-principles calculations can greatly improve the accuracy[10,11] of CALPHAD modeling. The present work develops a high throughput Density Functional Theory ToolKit (DFTTK) to perform DFT-based first-principles calculations at finite temperatures and variable compositions, based on the *atomate* package [12] from the *Materials Project* [5] and using the Vienna Ab initio Simulation Package (VASP) [13,14] as the calculation engine. The main benefits of *atomate* are its flexibility and data management platform, in particular the numerical convergence control, the numerical exception handling, and the usage of MongoDB database for DFT job control and DFT output data management. The properties at finite temperatures can be obtained both by the phonon approach based on lattice dynamics [15,16] and by the Debye–Grüneisen model [17,18], incorporating our extensive experience in theoretical methods[19–21], high-throughput calculations[22,23], structure generation[24,25], and software development [26–28]. To perform DFT-based first-principles calculations using *DFTTK*, the user needs only to name a structure file, namely, a VASP POSCAR file, either prepared by the user or produced by *DFTTK* by elemental substitution from a prototype structure.

The goals of DFTTK are to:

i. Automate file management and high-performance computing (HPC) job submission while providing clear, efficient and error-preventing default settings for high-throughput first-principles calculations;

ii. Establish robust relaxation schemes for handling both stable and unstable structures;

iii. Enable the development of workflows for temperature-dependent properties, starting with the free energies and its derivatives;

iv. Provide an interface for generating structures within the compound energy formalism (CEF) and exporting calculation data to formats compatible with CALPHAD modeling tools [27,29];

v. Create a free energy database for machine learning, data provenance and calculation reproducibility; and

vi. Automate the plotting of thermodynamic properties for analysis and publication.

## 2  Overview of methodologies for computing thermodynamic properties at finite temperatures and variable compositions

### 2.1  Helmholtz energy

It is known that the lattice parameters are more convenient input data for the majority of implementations of first-principles approaches. Therefore, it is more convenient to start from the Helmholtz energy [30–33]. According to the previous experiences [16,19,34], it is a reasonable approach to separate the Helmholtz energy into three accumulative contributions:

$$F(V,T) = E_c(V) + F_{vib}(V,T) + F_{el}(V,T) \qquad \text{Eq. 1}$$

where $F(V,T)$ represents the Helmholtz energy per atom at volume $V$ and temperature $T$; $E_c$ is the 0 K static total energy; $F_{vib}$ is the vibrational contribution; and $F_{el}$ is the thermal electronic contribution.

In *DFTTK*, one can choose to calculate the vibrational free energy of $F_{vib}$ in **Eq. 1** either by phonon approach based on lattice dynamics [15,16] or by Debye–Grüneisen model [17,18]. One is recommended to employ the phonon approach [19] because it is mathematically accurate in the second order and its input data can completely be predicted by DFT calculation without using any adjustable parameters. It is noted that with the modern computer, phonon calculations are straightforward for system with primitive unit cell containing less than 50 atoms if the space group symmetry of the system is relatively high. For extremely large system with low space group symmetry, such as those calculations based on special quasirandom structures (SQS) [24,35,36], one is recommended to use the primitive unit cell to perform phonon calculations. The Debye–Grüneisen model is implemented in *DFTTK* in case of when the structure is unstable or when the phonon calculations are too computationally demanding. It should be mentioned here that the Debye–Grüneisen model is only accurate at very low temperature and furthermore it also needs an empirical parameter to determine the Debye temperature [36].

## 2.2 Phonon approach to lattice vibration

The vibrational contribution to the Helmholtz energy by phonon theory can be computed by

$$F_{vib}(V,T) = k_B T \int_0^\infty \ln[\,2\sinh\frac{\hbar\omega}{2k_B T}]g(\omega,V)d\omega \qquad \text{Eq. 2}$$

where $k_B$ is the Boltzmann's constant, $\omega$ represents the phonon frequency, and $g(\omega,V)$ is the phonon density of states (PDOS).

## 2.3 Debye–Grüneisen model to lattice vibration

Strictly speaking, the formalism of the Debye–Grüneisen model [17,18] is only accurate at very low temperature of 10 K level [36,37]. It assumes a parabolic dependence of the PDOS on the phonon frequency, which is far from true for a realistic material as to be seen later in this study; at low temperatures, only the low frequency acoustic phonons are activated, which makes this relationship parabolic. That is why there are two classifications of Debye temperature: low and high. The low-temperature Debye temperature can be strictly derived by fitting low temperature heat capacity data. The high-temperature Debye temperature is mostly a phenomenological fitting parameter. The vibrational Helmholtz energy by the Debye approach is written as

$$F_{vib}(V,T) = \frac{9}{8}k_B\Theta_D(V) + k_B T\left\{3\ln\left[1-\exp\left(-\frac{\Theta_D(V)}{T}\right)\right] - D\left(\frac{\Theta_D(V)}{T}\right)\right\} \qquad \text{Eq. 3}$$

where $\Theta_D(V)$ is the Debye temperature and $D(x)$ is the Debye function given by

$$D(x) = 3/x^3 \int_0^x t^3/[exp(t) - 1]dt \qquad \text{Eq. 4}$$

$\Theta_D(V)$ can be calculated in through bulk modulus $B_0$ and Grüneisen constant $\gamma$ at the 0 K equilibrium atomic volume $V_0$.

$$\Theta_D(V) = \Theta_0 \left(\frac{V_0}{V}\right)^\gamma \qquad \text{Eq. 5}$$

where $\Theta_0$ represents the Debye temperature at the 0 K equilibrium atomic volume.

$$\Theta_0 = s(6\pi^2)^{1/3} \frac{\hbar}{k_B} V_0^{1/6} \left(\frac{B_0}{M}\right)^{1/2} \qquad \text{Eq. 6}$$

where $M$ is the average atomic mass and $s$ is a scaling factor, which calculated via the Poisson ratio $\upsilon$, as

$$s = 3^{5/6} \left[4\sqrt{2}\left(\frac{1+\upsilon}{1-2\upsilon}\right)^{3/2} + \left(\frac{1+\upsilon}{1-\upsilon}\right)^{3/2}\right]^{-1/3} \qquad \text{Eq. 7}$$

The Poisson ratio can be calculated if the elastic constants are known [38–40]. If the elastic constants are not known, the Poisson ratio can be approximately as a constant $\upsilon = 0.363615$ as suggested by Moruzzi et al. for cubic non-magnetic metals [17], resulting in $s = 0.617$. The Grüneisen constant can be calculated by [40]

$$\gamma = \frac{(1 + B_0')}{2} - x \qquad \text{Eq. 8}$$

Where $B_0'$ is the pressure derivative of the bulk modulus at the at the 0 K equilibrium volume with x = 2/3 for the case of high temperatures and x = 1 for the case of low temperature [17], or x = 4/3 by free-volume theory [18,41]. Table 1 lists the calculated $V_0$, $B_0$, $B_0'$, and $\Theta_0$ by DFT in this work for the compounds that will be used as the *DFTTK* examples discussed later in this work.

Table 1. Calculated $V_0$, $B_0$, $B_0'$, and $\Theta_0$ by DFT.

| Compound | $V_0$ (Å³/atom) | $B_0$ (GPa) | $B_0'$ | $\Theta_0$ (K) |
|---|---|---|---|---|
| Al | 16.4817 | 77.3165 | 4.676 | 384.9529 |
| Al$_3$Ni | 14.7434 | 112.3381 | 4.3492 | 400.4324 |
| Al$_4$Ni$_3$ | 13.3209 | 138.7464 | 4.3605 | 405.8721 |
| AlNi | 12.1261 | 158.185 | 4.3552 | 415.2047 |
| Al$_3$Ni$_5$ | 11.7534 | 166.8775 | 4.5004 | 405.8834 |
| AlNi$_3$ | 11.3761 | 180.3942 | 4.5715 | 402.9928 |
| Ni | 10.9297 | 196.5788 | 4.893 | 388.6386 |
| SQS-Al$_3$Ni | 14.29 | 109.1335 | 4.3684 | 392.6305 |
| SQS-AlNi | 12.6227 | 141.8267 | 4.4681 | 395.7887 |
| SQS-AlNi$_3$ | 11.5311 | 172.4344 | 4.6789 | 394.8915 |

## 2.4 Thermal electronic contribution

For the thermal electronic contribution, an acceptable approximation to calculate $F_{el}$ in **Eq. 1** is to use Mermin statistics [16,42] through

$$F_{el}(V,T) = E_{el}(V,T) - TS_{el}(V,T) \qquad \text{Eq. 9}$$

where the bare electronic entropy $S_{el}$ takes the form

$$S_{el}(V,T) = -k_B \int n(\varepsilon,V)\{f(\varepsilon,V,T)\ln f(\varepsilon,V,T) + [1 - f(\varepsilon,V,T)]\ln[1 - f(\varepsilon,V,T)]\}d\varepsilon \qquad \text{Eq. 10}$$

with $n(\varepsilon, V)$ being the electronic density-of-states as a function of one-electron band energy $\varepsilon$, $f$ is the Fermi function. The thermal electron energy takes the form

$$E_{el}(V,T) = \int n(\varepsilon, V) f(\varepsilon, V, T) \varepsilon d\varepsilon - \int^{\varepsilon_F} n(\varepsilon, V) \varepsilon d\varepsilon \qquad \text{Eq. 11}$$

where $\varepsilon_F$ is the Fermi energy

## 2.5 Quasiharmonic approximation (QHA)

In the *DFTTK* package, the QHA approach [36,43] is implemented by first calculating the Helmholtz energy in **Eq. 1** at several selected volumes (7 points by default following the previous experiences [19–22]) near the 0 K equilibrium volume, followed by the numerical interpolation using the 4-parameter Birch-Murnaghan equation of states (EOS) [40,44] to find the Helmholtz energy at an arbitrary volume. The default volume interval is set to 5% of the 0 K equilibrium volume. Too small volume interval might result in numerical instability due to the numerical uncertainties. Whenever available, *DFTTK* makes best use of the analytic formulas instead of the numerical second-order derivative to avoid numerical errors. For instance, when the phonon approach is employed, the constant volume heat capacity has the analytic expression in terms of phonon density of states [16].

## 2.6 Special quasirandom structures (SQS)

SQS [24,35,36] are specially small-unit-cell periodic supercell structures, designed to closely mimic the most relevant, near-neighbor pair and multisite correlation functions of random substitutional alloys. Generated based on an Ising lattice model, SQS can capture local relaxations. The correlation functions are classified by the cluster size (pair, triple, and quadruple) and the atomic distance within the cluster. As a result, the distribution of distinct local environments is maintained, and their average corresponds to the random alloy. Thus, DFT calculations using SQS can produce many important alloy properties that are dependent on the existence of those distinct local environments. The SQS approach has been used extensively to study the formation enthalpies, bond length distributions, density of states, band gaps and optical properties in semiconductor alloys [24].

The key quantities in the SQS approach are the *n*-site correlation functions. Describing random alloys by small unit-cell periodical structures could introduce erroneous correlations beyond a certain distance. However, since interactions between nearest neighbors are generally more important than interactions between more distant neighbors, SQS can be constructed in such a way that they reproduce the correlation functions for the first few nearest neighbors, deferring errors due to periodicity to more distant neighbors. The practical procedure [24] is to find the structures that match the 2-site correlation functions up to a given neighboring distance, and then add the conditions matching the high order correlation functions up to certain correlation distance. In this regard, a large database of pre-generated SQS that covers over 30 of the most common multi-sublattice structures has been provided by van de Walle et al. [45].

## 3    Compound energy formalism in the CALPHAD approach

The CALPHAD approach uses the compound energy formalism (CEF) for solution phases with sublattices[46,47]. The choice of sublattices is primarily in terms of the Wyckoff positions in the phases with each element entering into one or more sublattices. The DFT-based first-principles calculations have demonstrated remarkable agreement on energetics of solid phases with experimental data at their stable compositions [48] and thus provide valuable input data for CALPHAD modeling of individual phases [11,49]. *DFTTK* can thus further enhance the integration of data from the DFT-based first-principles calculations into the CALPHAD modeling.

There are usually more than one element mixing in a sublattice, which can be treated by the SQS approach discussed above. However, from the modeling point of view, the free energy of a phase with one element only in one sublattice is needed. For example, for random solution phases with one sublattice, the CEF indicates that one element can fully occupy the sublattice of the phase with the structure different from the ground state of the element, such as bcc Ni in the bcc Ni-Cr solution phase. The concept of lattice stability was thus introduced to denote the energy difference between the ground state and a non-ground state of the element [50]. The same concept applies to phases with multiple sublattices when each sublattice is occupied by one element only, denoted as endmembers of the phase, and the lattice stability of pure element is thus a special case of phases with one sublattice.

It is evident that the values of lattice stability are difficult to obtain from experiments, because the endmembers may not be stable, and have thus been estimated through extrapolations in the CALPHAD approach [7–9]. When the endmembers are metastable, their free energy can be reliably calculated using *DFTTK*. When the endmembers are unstable, there are imaginary phonon frequencies, and their entropies could thus not be directly calculated by DFT-based first-principles calculations as shown by **Eq. 2**. Even though our recently developed multiscale entropy approach is capable of predicting the instability [11,51–54], it is extremely computationally expensive, and more developments are needed for adoption in CALPHAD modeling. On the other hand, the enthalpies of unstable endmembers can be calculated by fixing the symmetry of their supercells, but values thus obtained are very different from the extrapolated values used in the CALPHAD modeling[23]. This is understandable because the extrapolations could not reach the unstable states with fixed symmetry. Van de Walle et al. recently showed that the values of the CALPHAD lattice stability are close to enthalpies at the limit of stability, i.e. the inflection point [55,56]. Yang et al. emphasized the importance for all endmembers to have the same symmetry in CALPHAD modeling and demonstrated that the adoption of symmetry-preserved first-principles calculations can serve the CALPHAD modeling well [57]. The relaxation scheme in *DFTTK* is designed similarly to preserve the symmetry of supercells as discussed below.

## 4   Description of the DFTTK package

With the thermodynamic properties at finite temperatures predicted by the QHA approach, other derived properties such as elastic constants [16] and thermoelectric properties [58–60] can be obtained using a quasi-static approach [16] according to the property-volume/strain relationship from the QHA approach.

### 4.1   *DFTTK* workflow

By default, DFTTK uses the VASP code [61] as the calculation engine through *pymatgen* [62] and *atomate* [12], and the user only needs to prepare the crystal structure file named POSCAR. For other parameters required by VASP, the default settings give a reasonable tradeoff between the calculation accuracy and computational time. If needed, the user can modify the default settings as shown by examples in Section 6, such as the $k$ point mesh density and the energy cutoff for structure relaxation. For example, the default energy cutoff of 520 eV and $k$ point mesh density of 8000/atom may be too high and thus costly for metallic system especially for SQS calculations. Based on the cross platform functions of the *pymatgen* [62] and *atomate* [12], the computational tasks can be first submitted to the MongoDB database from a computer and latterly run the tasks in other computers that have VASP installed and have access to the MongoDB database.

*DFTTK* is based on the workflow model in the *FireWorks* (FW) package developed by Jain et al. [63] with three main components: Firetask (an atomic computing job), Firework (JSON spec that includes all the information needed to bootstrap jobs), and Workflow (a set of FireWorks with dependencies between them). The workflows for *DFTTK* are schematically depicted in Figure 1. Managed by *Firetask* from *FireWorks* [12,63], the computational tasks are made of the structure relaxation (Robust Optimize/CheckRelax), energy-volume curve and QHA phonon calculations (EVcheck_QHA), and thermodynamic calculations (QHA Analysis).

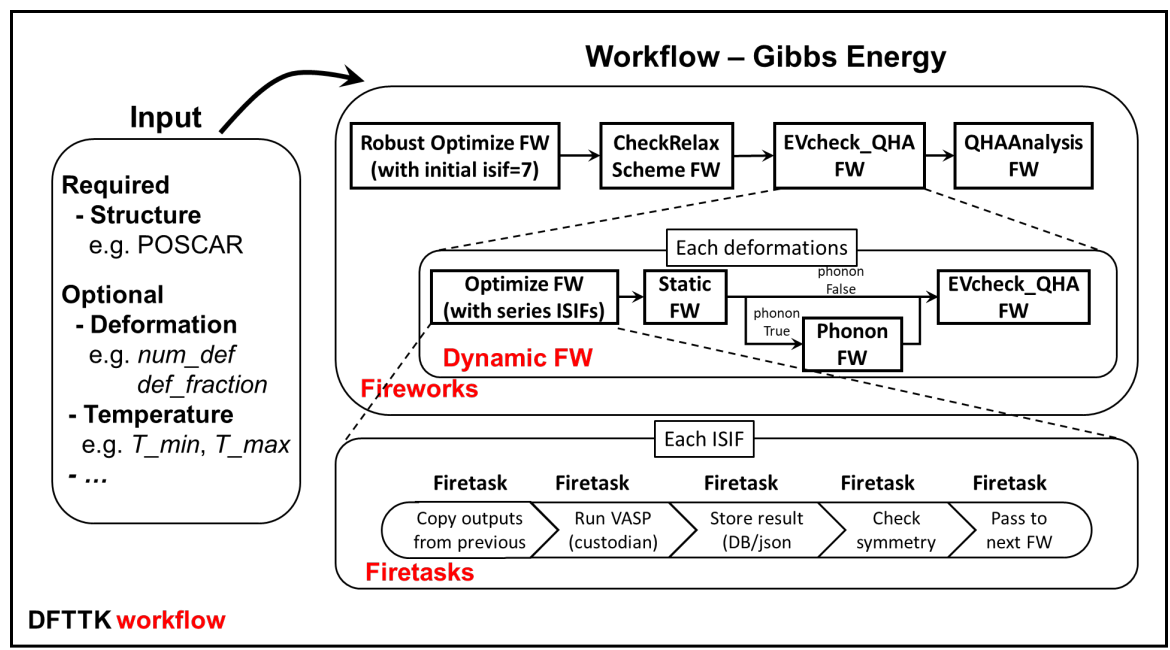

Figure 1. Workflows of *DFTTK* with FW [12,63] for the major computational steps such as the initial search for the equilibrium volume, lattice shape relaxation, static calculation, phonon calculation, and QHA analysis, managed by Firetask [12]. ISIF (see Table 2) is the VASP settings [13,14] controlling the various structure relaxation scheme.

## 4.2 Relaxation scheme of *DFTTK*

A key feature of *DFTTK* is the structure relaxation scheme under the loose constrain conditions to guarantee that the symmetry of the relaxed structure is not seriously distorted from that of the initial structure. This information is especially useful in CALPHAD modeling [27,28] on calculation of the energetics for structures that may be unstable at 0 K, for instance

1. Endmembers generated by the compound energy formalism;
2. High temperature phases; and
3. Random solution phase under SQS approach [24,35,36].

Table 2. Structure relaxation scheme of VASP [13,14]

| ISIF | calc. force | calculate stress tensor | relax ions | change cell shape | change cell volume |
|---|---|---|---|---|---|
| 0 | yes | no | yes | no | no |
| 1 | yes | trace only | yes | no | no |
| 2 | yes | yes | yes | no | no |
| 3 | yes | yes | yes | yes | yes |
| 4 | yes | yes | yes | yes | no |
| 5 | yes | yes | no | yes | no |
| 6 | yes | yes | no | yes | yes |
| 7 | yes | yes | no | no | yes |

The solution is to make the proper use of ISIF control parameter in VASP [13,14], which controls structure relaxation scheme, by means of a variety of relaxation combinations among atomic volume, lattice shape, and atomic positions, given in Table 2. *DFTTK* makes the combinations of the ISIF options in Table 2 optimized by going through calculation steps by the flow chart illustrated in Figure 2 where "Pass or Passed" means that the final relaxed structure is not seriously distorted from the initial structure (judged by the given threshold values such as lattice shape change from ideal lattice or atomic displacement from ideal position, see references by van de Walle et al. [24,43]), whereas "Fail" means otherwise. There are two relaxation levels as shown in Figure 2 with the difference discussed below. The default setting in DFTTK is the Level 1, and its steps are briefly presented as follows:

- Step 1: Fast finding the rough initial equilibrium cell volume with ISIF=7 with only cell volume change. Skip to Step 3 by default, or continue to Step 2 by user option.

- Step 2: Further relax the lattice shape and atomic positions. If this step passes, continue to the QHA calculations using the phonon approach or Debye model to calculate thermodynamic properties. Otherwise, go to Step 3;
- Step 3: Only relax lattice shape based on the relaxed structure from Step 1. If this step passes, go to Step 4. Otherwise, go to Step 5;
- Step 4: Further relax the lattice shape and atomic positions. If this step passes, continue to QHA calculations using phonon approach or Debye model to calculate thermodynamic properties. If this step fails, go to Step 5;
- Step 5: Only relax atomic positions based on the relaxed structure from Step 1. If this step fails and Step 3 passes, use the result from Step 3 and continue to QHA calculations using Debye model to calculate thermodynamic properties. If this step fails, calculate the energy without lattice relaxation based on the relaxed structure from Step 1 and continue to QHA calculations using the Debye approach to calculate thermodynamic properties. If this step passes, go to Step 6;
- Step 6: Continue to relax the structure according to the predetermined relaxation level given by the user. For the default relaxation level 1, check if Step 3 passes. If Step 3 passes, use low energy one between Step 3 and Step 5 and continue to QHA calculations using Debye model to calculate thermodynamic properties. Otherwise, use the energy from Step 5 and continue to QHA calculations using Debye model to calculate thermodynamic properties. For the relaxation level 2, further relax the lattice shape and atomic positions. If this step passes, use the energy and continue to QHA calculations using Debye model to calculate thermodynamic properties. Otherwise, if Step 3 passes, use low energy one between Step 3 and Step 5 and continue to QHA calculations using

Debye model to calculate thermodynamic properties.; if Step 3 fails, use the energy from Step 5 and continue to QHA calculations using Debye model to calculate thermodynamic properties.

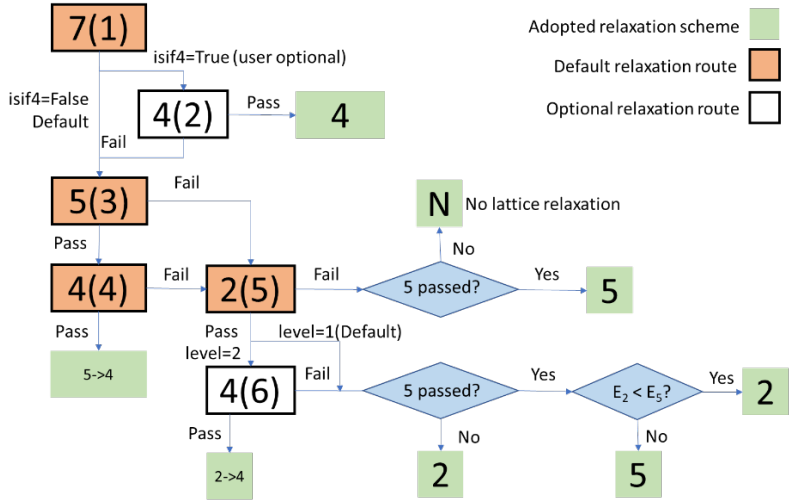

Figure 2. Relaxation scheme of *DFTTK* based on the VASP ISIF setting numbers given in Table 2 and the steps (number in parenthesis) described in the text. The relaxation is initiated by the fast rough searching for the equilibrium volume with ISIF=7 followed by the various branches marked by the ISIF values (2-7) indicated in the squares according to various criteria of user options. Green color means the final adopted scheme and "N" marked in the square shape means no lattice relaxation will be performed.

# 5  Installation and run *DFTTK*

*DFTTK* [64] is distributed on the Python package index (PyPI) with Python 3.6 and above under the name "dfttk". To install *DFTTK*, one can simply run *pip install dfttk*. *DFTTK* uses MongoDB to manage/store job tasks, inputs, and outputs for calculations, and use *pymatgen* to generate some inputs for running VASP. Hence, the user should use *dfttk config [option]* to configure the MongoDB and *pymatgen* before running calculations. The details run/configure DFTTK are detailed in *DFTTK* online documentation [64]. To run *DFTTK*, the user should use the following command style:

```
dfttk module [option]
```

where the available modules are given in Table 3. The user can use *dfttk -h* to get all the available modules and use *dfttk module -h* to get the detail options for each module.

Table 3. Available modules in *DFTTK*

| module | function |
|---|---|
| *run* | DFT job submission. Create workflows which call VASP to perform 0 K E-V, phonon, elastic moduli, Born effective charge (dielectric tensor), etc |
| *config* | Set basic *DFTTK* run environment, such as path to pseudopotentials and batch job submission templates |
| *db_remove* | Remove certain data in MongoDB to save storage cost |
| *thelec* | Download data from MongoDB database & postprocess them to get thermodynamic properties, elastic constant, etc. |
| *thfind* | Check the MongoDB database for DFT results followed by calling the *thelec* module to get thermodynamic properties when the option -get is given. |
| *EVfind* | Find the metadata tags that have 0 K static calculation finished. |

# 6 Calculations examples

The examples are designed for users to test the *DFTTK* package, currently including i) the stoichiometric phases of Al, Ni, $Al_3Ni$, AlNi, $AlNi_3$, $Al_3Ni_4$, and $Al_3Ni_5$, and ii) the fcc solution phases of $Al_3Ni$, AlNi, $AlNi_3$ treated using SQS. The examples can be downloaded from a separate GitHub repository [65]. In the following subsections, we will brief the calculations for stoichiometric phases of Al, Ni, AlNi, and $AlNi_3$ and the fcc solution phases of $Al_3Ni$, AlNi, $AlNi_3$. For the interested readers, the calculational setting for the stoichiometric phases of $Al_3Ni$, $Al_3Ni_4$, and $Al_3Ni_5$ can be downloaded from the example repository [65].

## 6.1 Al example

The Al example comes with two input files. The one named "POSCAR" is the regular VASP POSCAR file based on the primitive unit cell as follows

```
Al1
4.10
0 .5 .5
.5 0 .5
.5 .5 0
Al
1
direct
0 0 0 Al
```

The one named "SETTINGS.yaml" is made of the following two lines to instruct *DFTTK* to perform QHA calculations using the small displacement method [66].

```
phonon : True
phonon_supercell_matrix : [[-3, 3, 3], [3, -3, 3], [3, 3, -3]]
```

where the supercell transformation matrix of `[[-3, 3, 3], [3, -3, 3], [3, 3, -3]]` for the phonon calculation results in a 3×3×3 cubic supercell of the conventional cubic unit cell for fcc structure. Using cubic supercell for cubic system is recommended for phonon calculation [67] due to the consideration to keep the cubic symmetry of the calculated force constants.

The command to run the Al example is

```
dfttk run -wf robust -f POSCAR -l -m 1
```

where the options are: "`run`" to instruct *DFTTK* to execute the "`run`" module, "`-wf robust`" to perform robust structural optimization as detailed in Section 4.2, "`-f POSCAR`" to tell the structure file name is "`POSCAR`", "`-l`" to schedule the task to the MongDB database, and "`-m 1`" to submit a batch job into the computer. If one wants to submit a *DFTTK* task using a POSCAR file stored in a computer without VASP installed, the command should be split into first to run "`dfttk run -wf robust -f POSCAR -l`" followed launching the DFT batch job in the computers that have *DFTTK* and VASP installed using the *atomate* command "`qlaunch singleshot`". After these, one can check the progress of the calculations by the *atomate* command '`lpad get_wflows`' at any computers that have connection to the MongDB database. The completion of a specific *DFTTK* calculation is indicated when all the values for the 'states_list' fields are shown as 'C' in the output of the *atomate* command '`lpad get_wflows`'. After the task is completed, one can retrieve and plot the thermodynamic properties by running

```
dfttk thfind -get -plot DFTTK -fitF -expt ExptData.json
```

where the file ExptData.json under the dfttk_example folder contains experimental data for comparison with the *DFTTK* calculations. The above command will produce a folder named `Al_Fm-3m_225PBE/` which contains all calculated thermodynamic properties. In particular, it includes

- `figures` - plots in png format for most of the thermodynamic prosperities

- `readme` - extensive summary of the calculated results in JSON format

- `fvib_ele` - tabulated data containing the calculated thermodynamic properties

- `fvib_eij` - tabulated data containing the calculated thermal expansion coefficient tensor

- `record.json` - SGTE fitting record for heat capacity, Gibbs energy, enthalpy, and entropy at given temperature range

The calculated PDOS for Al at the 300 K theoretical equilibrium volume is illustrated in Figure 3 which agrees well the recent inelastic neutron scattering data [68] and previous calculation [69].

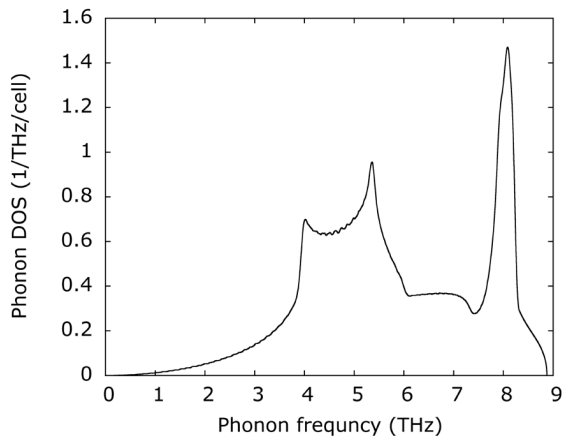

Figure 3. Calculated phonon density of states for Al.

The calculated linear thermal expansion coefficient for Al is compared with the measured data by Touloukian et al. [70] in Figure 4.

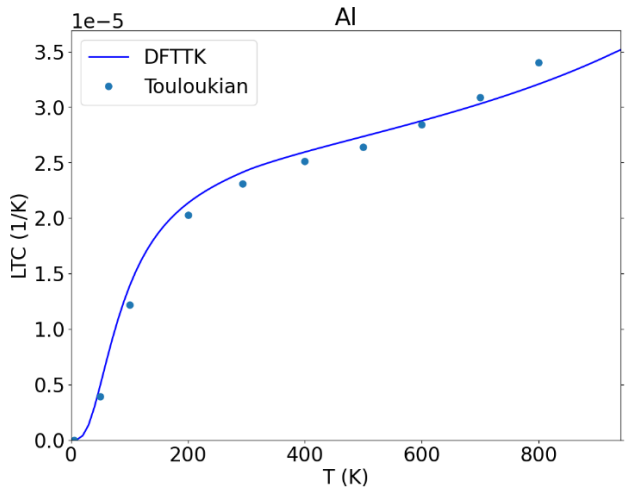

Figure 4. Thermal expansion coefficient of Al. The solid line represents the calculated result by *DFTTK* and the symbols represent the compiled experimental data by Touloukian et al. [70].

The calculated heat capacity, entropy, and enthalpy for Al are compared with the measured data [71,72], data from JANAF table [73], and data from CALPHAD database (SSUB) [74].. in Figure 5, Figure 6, and Figure 7, respectively.

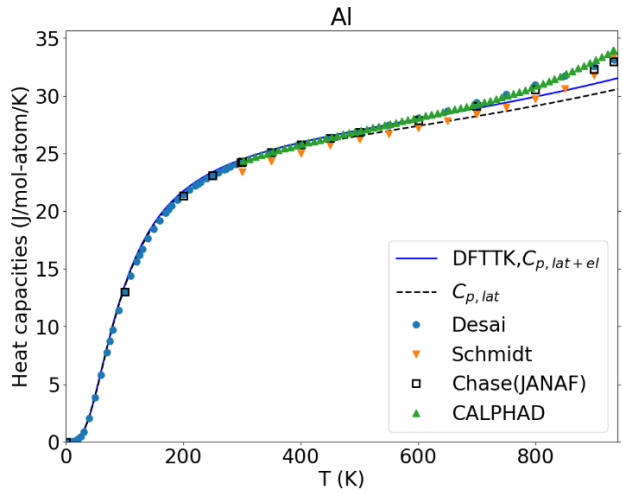

Figure 5. Heat capacity of Al. The solid and dashed lines represent the calculated results by *DFTTK* with and without considering the electronic contributions, respectively. The symbols represent for the measured data [71,72], data from JANAF table [73], and data from CALPHAD database (SSUB) [74].

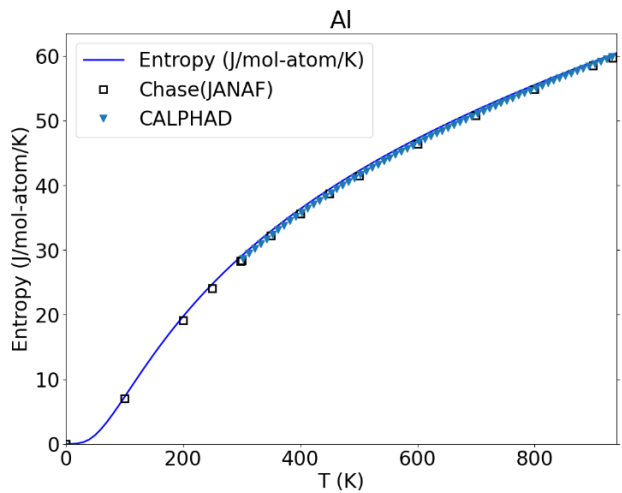

Figure 6. Entropies of Al. The solid line represents the calculated results by *DFTTK* and the symbols represent data from JANAF table [73] and data from CALPHAD database (SSUB) [74].

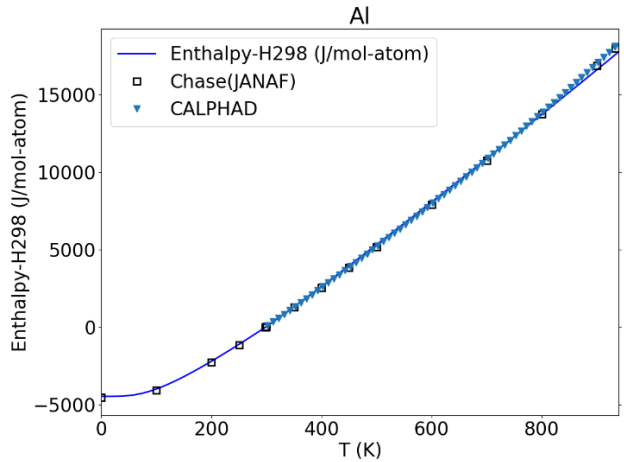

Figure 7. Enthalpies of Al. The solid line represents the calculated results by *DFTTK* and the symbols represent data from JANAF table [73] and data from CALPHAD database (SSUB) [74].

## 6.2 Ni example

The settings for Ni are quite similar to Al by replacing Al with Ni in the files shown in the previous section, except treated as ferromagnetic by default with the following lines in the "SETTINGS.yaml" file.

```
override_default_vasp_params :
    user_incar_settings :
        magmom : [1]
phonon_supercell_matrix : [[-3, 3, 3], [3, -3, 3], [3, 3, -3]]
```

where the key "magmom" is for setting the initial magnetic moment for Ni.

The calculated PDOS for Ni at the 300 K theoretical equilibrium volume is illustrated in Figure 8 which agrees well the recent inelastic neutron scattering data [75] and previous calculation [76].

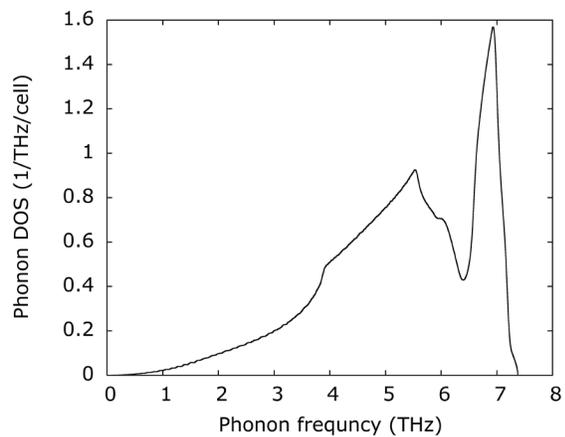

Figure 8. Calculated phonon density of states for Ni.

The calculated linear thermal expansion coefficient, heat capacity, entropy, and enthalpy for Ni are compared with the measured and modeled data [70,73,74,77–80] in Figure 9.

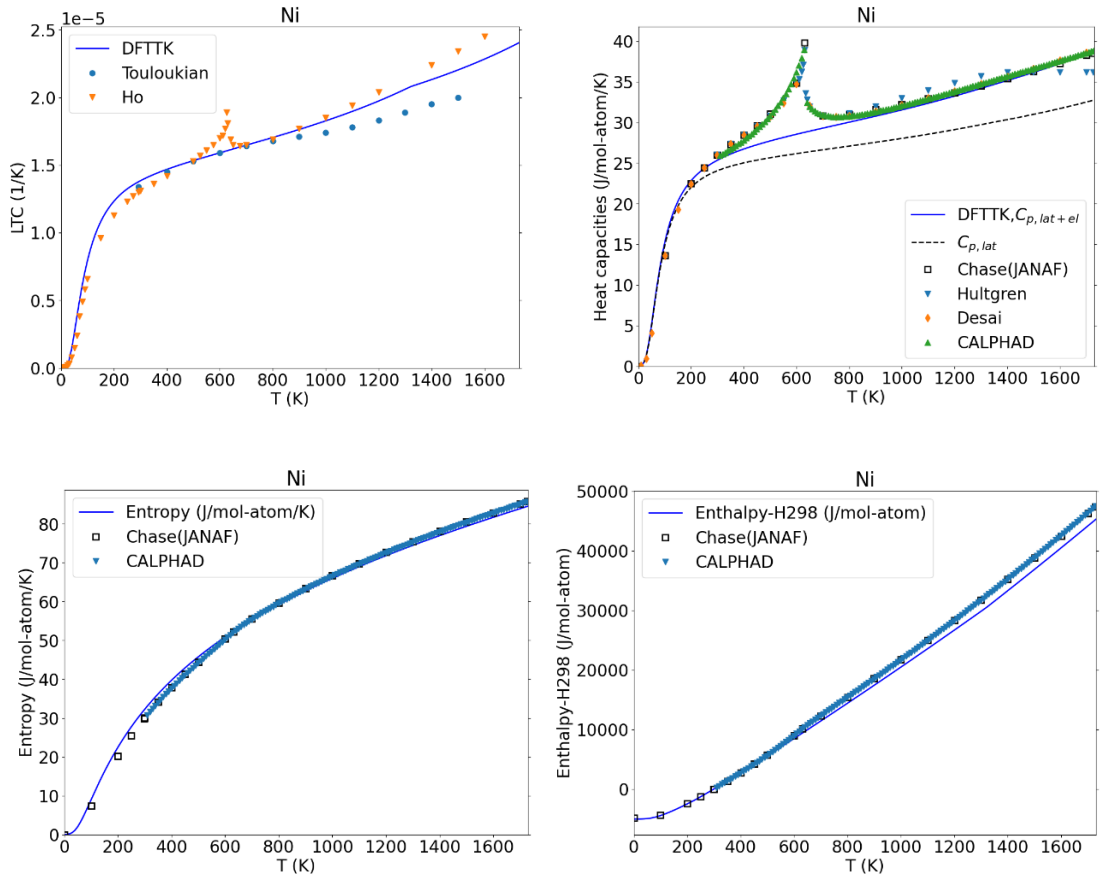

Figure 9. Linear thermal expansion coefficient, heat capacity, entropy, and enthalpy of Ni calculated by DFTTK in comparison with data from literatures [70,73,74,77,79–81].

6.3 Stoichiometric AlNi and AlN$_3$ example

The stoichiometric compounds of B2-AlNi and L1$_2$-AlNi$_3$ in the Al-Ni binary system are selected to demonstrate the *DFTTK* calculation for compounds. The content of the setting file (SETTINGS.yaml) is

```
phonon : True
phonon_supercell_matrix : [[2,0,0], [0,2,0], [0,0,2]]
```

Figure 10 shows the calculated phonon density of states for B2-AlNi and L1$_2$-AlNi$_3$. Figure 11 compares the calculated heat capacities with the experimental data or CALPHAD modeling [71,74,82–85], and Figure 12 compares their calculated entropies with the available CALPHAD modeling [74].

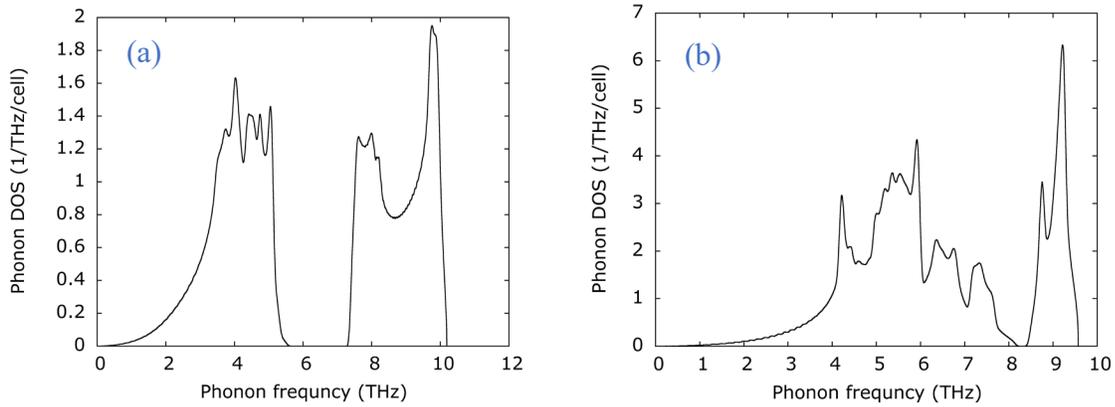

Figure 10. Calculated phonon density of states for stoichiometric compounds (a) B2-AlNi and (b) L1$_2$-AlNi$_3$.

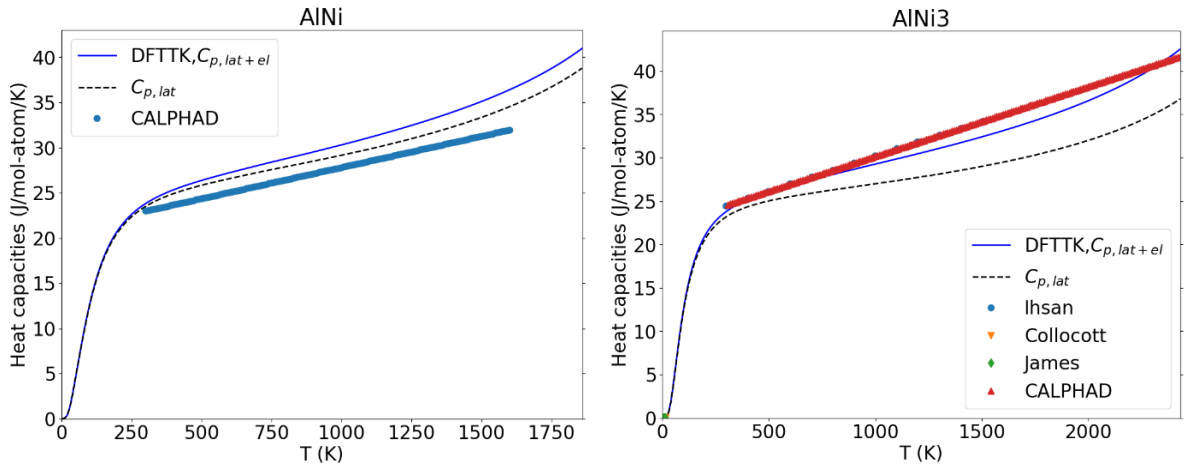

Figure 11. Heat capacities of B2-AlNi and L1$_2$-AlNi$_3$ with the solid and dashed lines representing the calculated results by *DFTTK* with and without considering the electronic contributions, respectively. The symbols represent the literature data [71,74,82–85] (experimental data reported by Sandakova et al. and Kucherenko and Troshkina are read from Miracle [85])

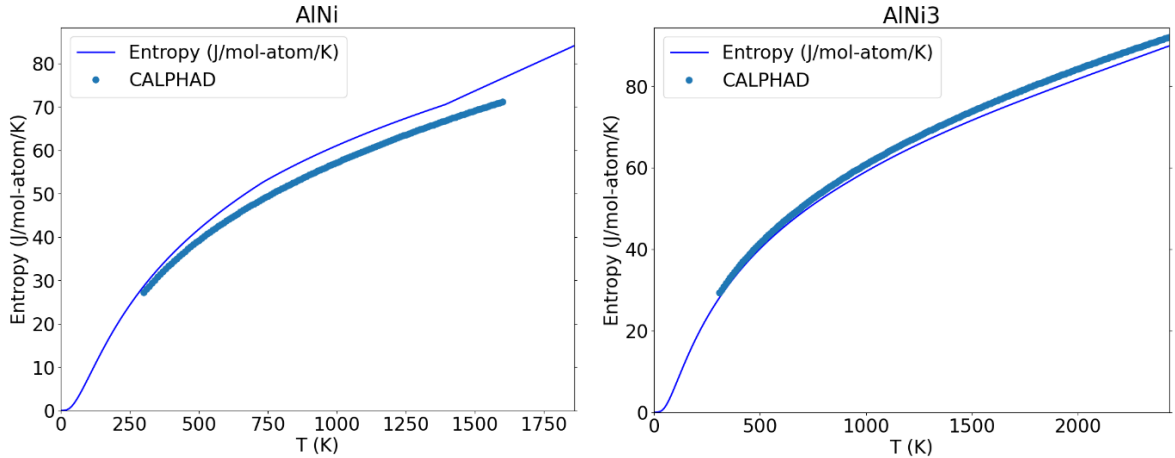

Figure 12. Entropies of B2-AlNi and L1$_2$-AlNi$_3$ with the solid lines for calculated results by *DFTTK* and the symbols for the CALPHAD data [74] for L1$_2$-AlNi$_3$.

### 6.4 SQS example

The SQS calculations were performed at three compositions of Al$_3$Ni, AlNi, and AlNi$_3$ for the fcc random solution. The initial POSCAR files based on 16-atom supercell prepared by this work are given in the Appendix A, B, and C, for the compositions Al$_3$Ni, AlNi, and AlNi$_3$, respectively. The content of the setting file (SETTINGS.yaml) is listed below

```
override_default_vasp_params :
```

```
    user_incar_settings :

        magmom: [16*1]

        Relax_settings:

            PREC: High

            grid_density: 1000

force_phonon : True

phonon : True

phonon_supercell_matrix : [[-1,1,1], [1,-1,1], [1,1,-1]]
```

where the keys of '`magmom`' are used to set the initial magnetic moments for the individual atoms, '`PRES`' and '`grid_density`' to use the default VASP setting for energy cutoff together with a reciprocal *k*-mesh density of 1000/atom to reduce the computational cost, '`force_phonon`' is to force *DFTTK* to perform phonon calculations even the relaxed structure is heavily distorted from the ideal structure or even the relaxed structure is not stable. Since the phonon calculations are time consuming, the super used cell is just a doubling of the primitive unit cell.

Figure 13 shows the calculated phonon density of states of fcc Al-Ni solution phases at compositions of $Al_3Ni$, AlNi, and $AlNi_3$. No imaginary phonon modes are found, demonstrating that the relaxed structures are stable. Figure 14 shows the calculated heat capacities of fcc Al-Ni solution phases at compositions of $Al_3Ni$, AlNi, and $AlNi_3$, and Figure 15 shows the calculated entropies.

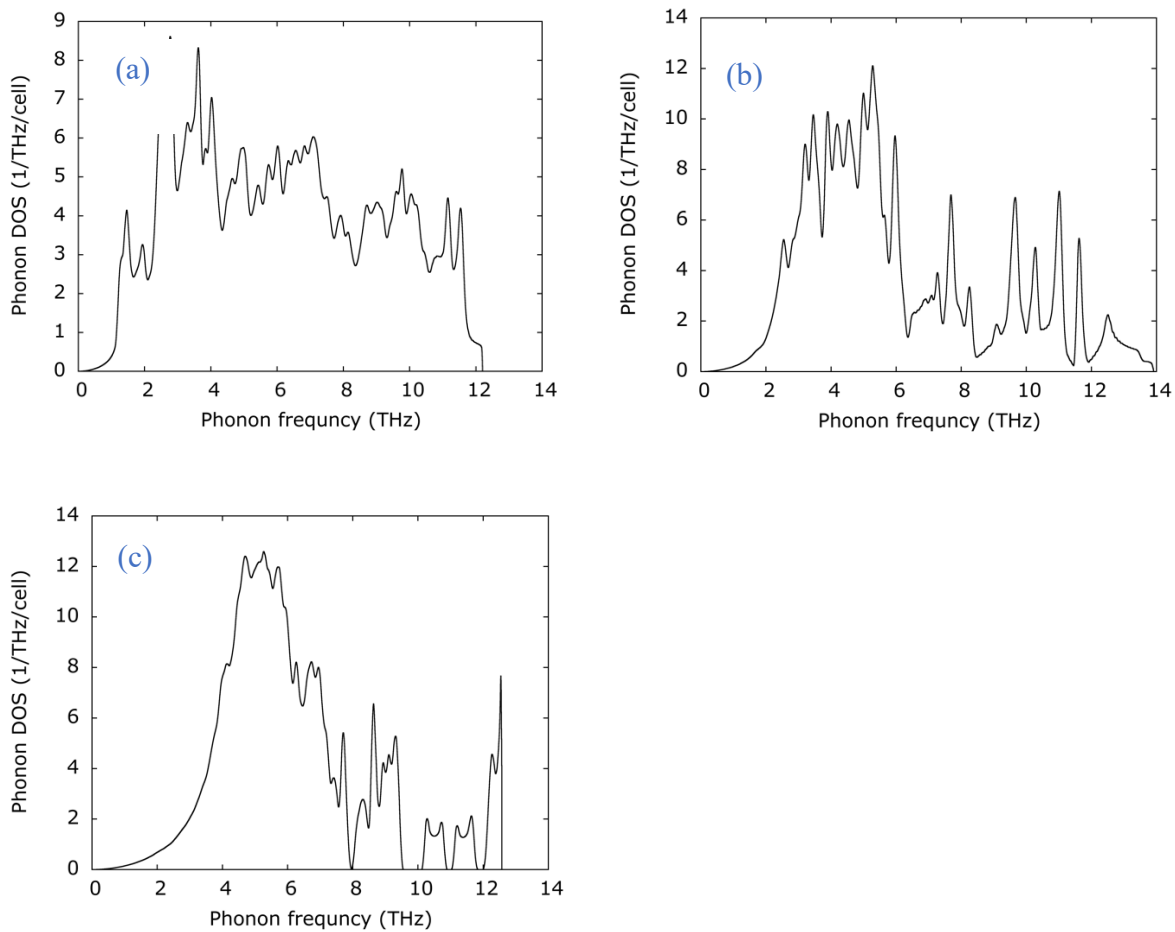

Figure 13. Calculated phonon density of states for the fcc Al-Ni solid solution phases at the compositions of (a) $Al_3Ni$, (b) AlNi, and (c) $AlNi_3$.

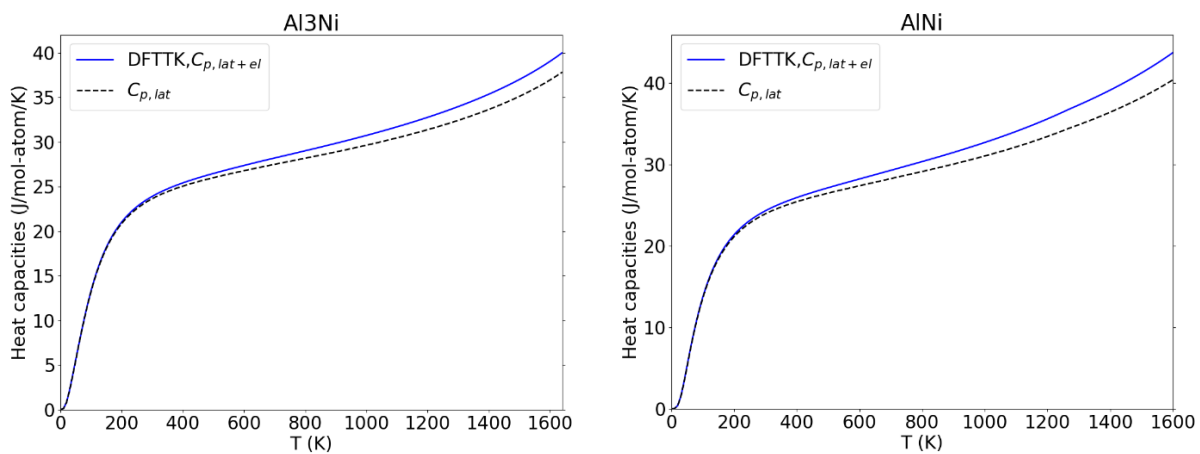

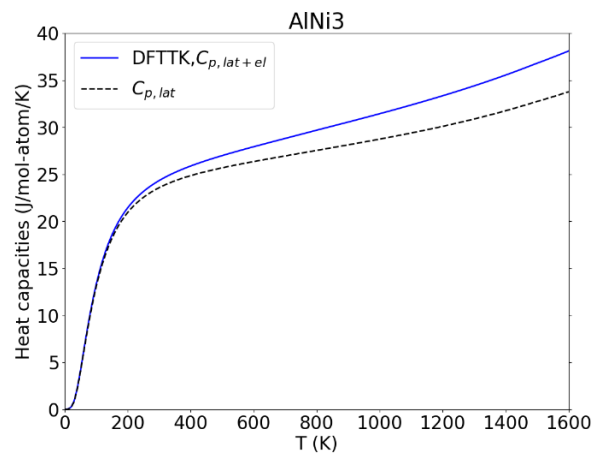

Figure 14. Heat capacities of the fcc Al-Ni solid solution phases at the compositions of Al$_3$Ni, AlNi, and AlNi$_3$ with the solid and dashed lines representing the calculated results by *DFTTK* with and without considering the electronic contributions, respectively.

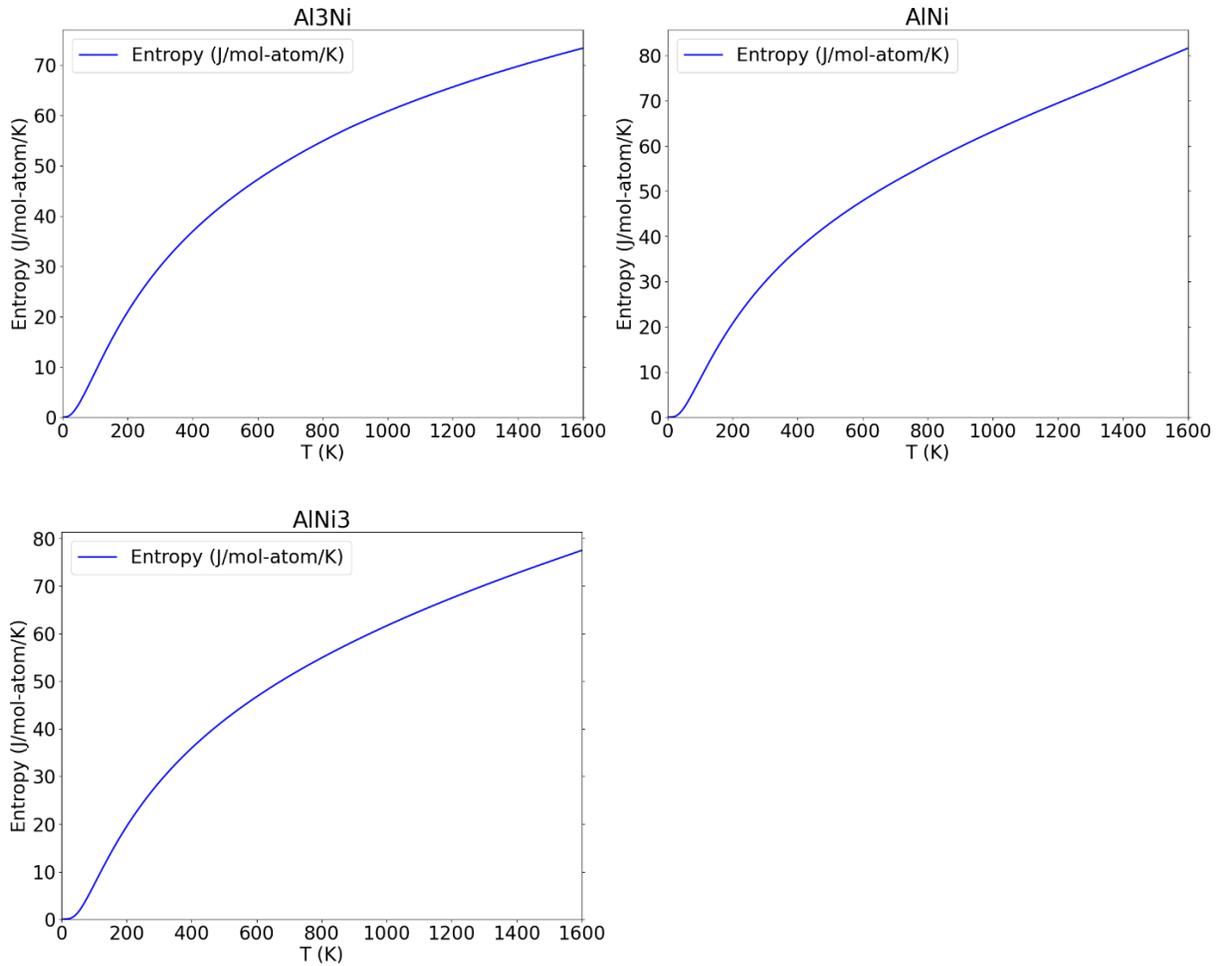

Figure 15. Calculated entropies for the fcc Al-Ni solid solution phase at the compositions of Al$_3$Ni, AlNi, and AlNi$_3$.

6.5   Example with Debye–Grüneisen model

This subsection briefs the calculated results with Debye–Grüneisen model using Al as an example treated at high temperature case, i.e., x = 2/3 in **Eq. 8**. *DFTTK* always performs the additional thermodynamic calculations using the Debye–Grüneisen model as long as the energy volume data are calculated. To check the results calculated using the Debye–Grüneisen model, one needs to run the following *DFTTK* command

```
dfttk thfind -get -renew -qhamode debye -all Al -plot DFTTK
-fitF -expt ExptData.json
```

The results are compared with experimental data [70,72,74,86,87] in Figure 16

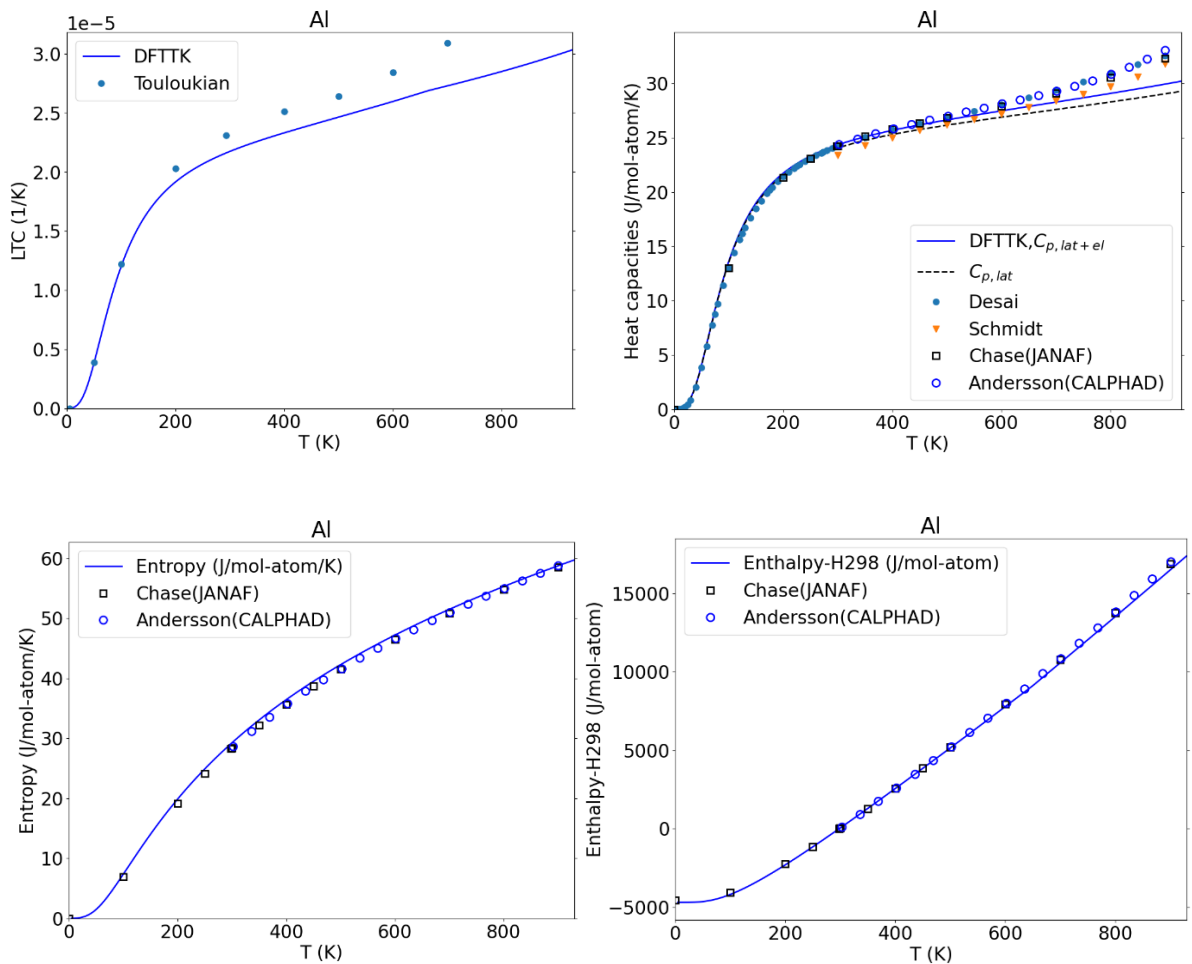

Figure 16. DFTTK results using Debye–Grüneisen model with Al as an example.

## 6.6 Plots of PDOS and phonon dispersions

If the Yphon package [26] is installed (see DFTTK documentation [64] on how to install Yphon), the PDOS and phonon dispersions can be automatically plotted at the 300 K theoretical equilibrium volume by an interpolation among the force constants obtained by the quasiharmonic calculations with the following *DFTTK* command

```
dfttk thfind -get -renew -py -w AlNi -plot DFTTK
```

The phonon dispersions for the stoichiometric phases of Al, Ni, AlNi, and AlNi3 are illustrated in Figure 17 and the phonon dispersions for the fcc solution phases of Al3Ni, AlNi, AlNi3. are illustrated in Figure 18.

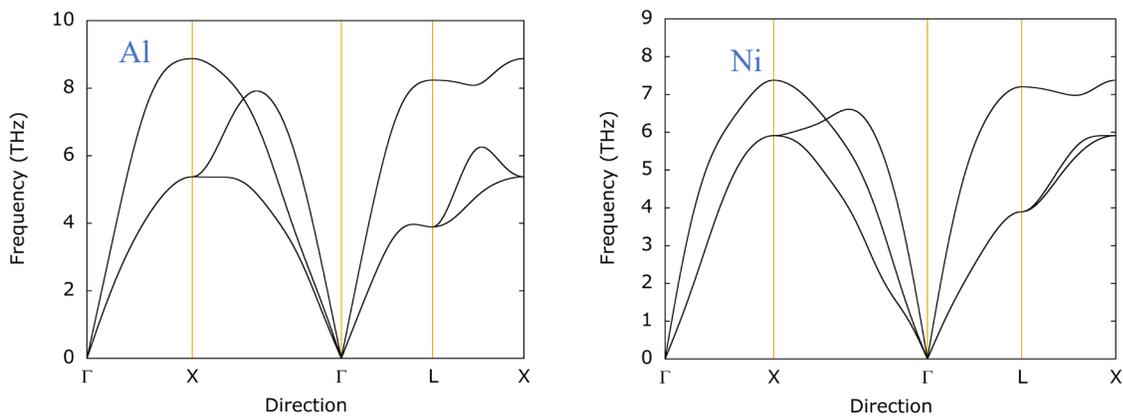

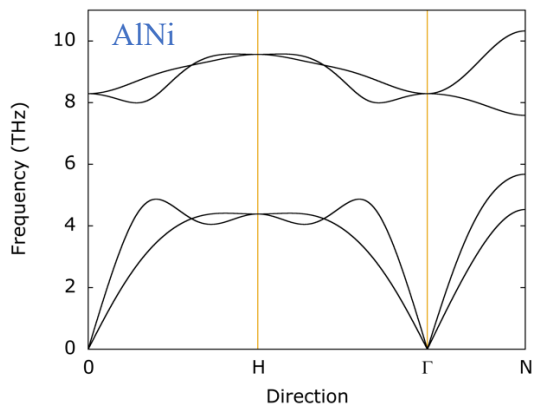
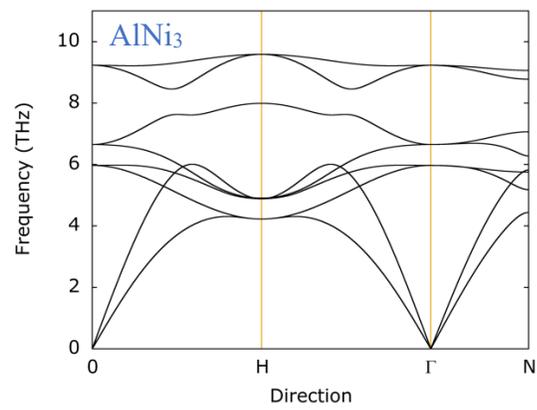

Figure 17. Calculated phonon dispersions for the stoichiometric phases of Al, Ni, AlNi, and AlNi$_3$.

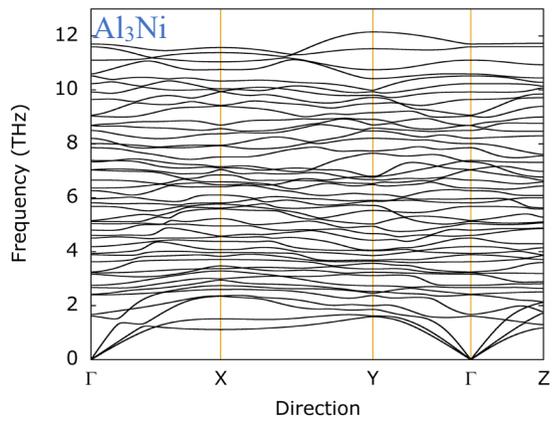
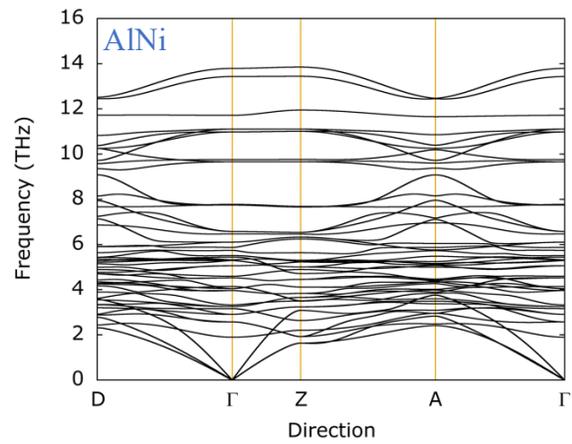

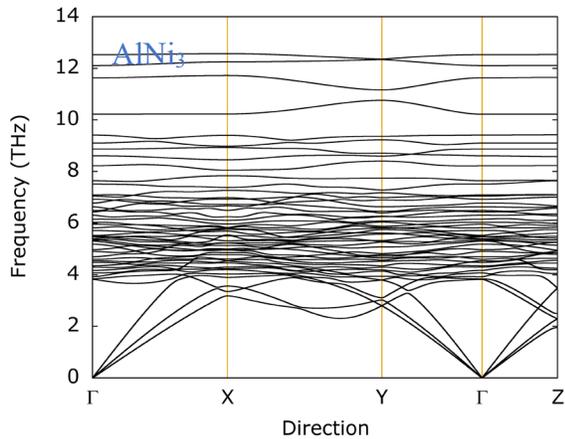

Figure 18. Calculated phonon dispersions phonon dispersions for the fcc solution phases of Al$_3$Ni, AlNi, and AlNi$_3$.

## 7 Summary

In summary, we have developed the open-source *DFTTK* package which integrates our accumulated experience in the high-throughput first-principles calculations of thermodynamic properties of materials at finite temperature either by phonon approach based on lattice dynamics [15,16] or by Debye–Grüneisen model [17,18]. The features of *DFTTK* include i) High-throughput for many structures with one simple command; ii) Simple settings with only the structure file required; iii) High-throughput post-processing of data stored in MongoDB with one simple command; iv) Search of the MongoDB database for existing calculation results and settings used; v) extraction of data from and restart of partially finished calculations; vi) Consideration of the thermal electron contribution to the thermodynamic properties; vii) Automated plotting of most of common thermodynamic properties, phonon dispersion/DOS with publication quality.

Executing the cross platform features of the *pymatgen* [62] and *atomate* [12] packages, the DFTTK package, written in Python, can be installed in major computer operating systems, such as Window, MacOS, and Linux. Examples of functionality are given using the Ni-Al system, being calculated for seven stoichiometric phases of Al, Ni, $Al_3Ni$, AlNi, $AlNi_3$, $Al_3Ni_4$, and $Al_3Ni_5$, and the fcc solution phase at the compositions of $Al_3Ni$, AlNi, and $AlNi_3$.

**Acknowledgments**

The authors at Penn State acknowledge the financial supports partially by the Department of Energy (DOE) via Award Nos. DE-AR0001435, DE-EE0008456, DE-FE0031553, DE-NE0008757, DE-NE0008945, and DE-SC0020147; partially by the Computational Materials Sciences Program funded by the DOE, Office of Science, Basic Energy Sciences, under Award Number DE-SC0020145 (Wang and Chen); partially by the National Science Foundation (NSF) with Grant Nos. CMMI-1825538 and CMMI-2050069; and partially by the Office of Naval Research (ONR) via Contract Nos. N00014-17-1-2567 and N00014-21-1-2608; and partially supported by a NASA Space Technology Research Fellowship Grant 80NSSC18K1168 (Bocklund and Liu). First-principles calculations were performed partially on the Roar supercomputer at the Pennsylvania State University's Institute for Computational and Data Sciences (ICDS), partially on the resources of the National Energy Research Scientific Computing Center (NERSC) supported by the DOE Office of Science User Facility operated under Contract No. DE-AC02-05CH11231, and partially on the resources of the Extreme Science and Engineering Discovery Environment (XSEDE) supported by NSF with Grant No. ACI-1548562.

**Data availability statement**

The raw data required to reproduce this work are available to download from [DOI: 10.5281/zenodo.5517543](DOI:10.5281/zenodo.5517543).

Appendix A. POSCAR file for fcc Al-Ni solution phases at the composition of $Al_3Ni$

```
3.55
    0.500000     0.500000   -1.000000
    0.000000    -1.500000   -0.500000
   -2.000000     0.500000   -0.500000
Ni Al
12 4
D
0.25000000 0.18750000 0.31250000
0.25000000 0.43750000 0.06250000
0.50000000 0.12500000 0.87500000
0.50000000 0.37500000 0.62500000
0.50000000 0.62500000 0.37500000
0.50000000 0.87500000 0.12500000
0.75000000 0.06250000 0.43750000
0.75000000 0.31250000 0.18750000
0.75000000 0.56250000 0.93750000
0.75000000 0.81250000 0.68750000
1.00000000 0.75000000 0.25000000
1.00000000 1.00000000 1.00000000
0.25000000 0.68750000 0.81250000
0.25000000 0.93750000 0.56250000
1.00000000 0.25000000 0.75000000
1.00000000 0.50000000 0.50000000
```

Appendix B. POSCAR file for fcc Al-Ni solution phases at the composition of AlNi

```
AlNi
1.00
        -3.5500000000         3.5500000000         7.1000000000
        -3.5500000000         7.1000000000         3.5500000000
        -7.1000000000         3.5500000000         3.5500000000
 Al Ni
 8 8
D
   0.2500000000    0.2500000000    0.2500000000
   0.2500000000    0.2500000000    0.7500000000
   0.2500000000    0.7500000000    0.2500000000
   0.5000000000    0.0000000000    0.0000000000
   0.7500000000    0.2500000000    0.2500000000
   0.0000000000    0.5000000000    0.0000000000
   0.0000000000    0.0000000000    0.5000000000
   0.0000000000    0.0000000000    0.0000000000
   0.2500000000    0.7500000000    0.7500000000
   0.5000000000    0.5000000000    0.5000000000
   0.5000000000    0.5000000000    0.0000000000
   0.5000000000    0.0000000000    0.5000000000
   0.7500000000    0.2500000000    0.7500000000
   0.7500000000    0.7500000000    0.2500000000
```

|    0.7500000000    |    0.7500000000    |    0.7500000000    |
|    0.0000000000    |    0.5000000000    |    0.5000000000    |

Appendix C. POSCAR file for fcc Al-Ni solution phases at the composition of AlNi$_3$.

AlNi3

1.00

```
        1.7750000000            1.7750000000           -3.5500000000
        0.0000000000           -5.3250000000           -1.7750000000
       -7.1000000000            1.7750000000           -1.7750000000
```

 Ni Al

 12 4

D

```
   0.2500000000     0.1875000000     0.3125000000
   0.2500000000     0.4375000000     0.0625000000
   0.5000000000     0.1250000000     0.8750000000
   0.5000000000     0.3750000000     0.6250000000
   0.5000000000     0.6250000000     0.3750000000
   0.5000000000     0.8750000000     0.1250000000
   0.7500000000     0.0625000000     0.4375000000
   0.7500000000     0.3125000000     0.1875000000
   0.7500000000     0.5625000000     0.9375000000
   0.7500000000     0.8125000000     0.6875000000
   0.0000000000     0.7500000000     0.2500000000
   0.0000000000     0.0000000000     0.0000000000
```

| | | |
|---|---|---|
| 0.2500000000 | 0.6875000000 | 0.8125000000 |
| 0.2500000000 | 0.9375000000 | 0.5625000000 |
| 0.0000000000 | 0.2500000000 | 0.7500000000 |
| 0.0000000000 | 0.5000000000 | 0.5000000000 |